\documentclass[pra,aps,superscriptaddress,showpacs,reprint,longbibliography,nofootinbib]{revtex4-1}

\usepackage{lmodern}
\usepackage[T1]{fontenc}
\usepackage{amsmath}
\usepackage{color}
\usepackage{amsfonts}
\usepackage{graphicx}
\usepackage[normalem]{ulem}
\usepackage{hyperref}
\hypersetup{
colorlinks=true,final=true,
        linkcolor=black,
        citecolor=black,
        filecolor=black,
        urlcolor=black,
        pdfauthor={},
        pdftitle={}
}
\usepackage[utf8]{inputenc}
\usepackage{amsmath,amssymb,amsfonts}
\usepackage{xcolor}

\newcommand{\braket}[1]{\left< #1 \right>}
\newcommand{\ketbra}[1]{\left| #1 \right>\left< #1 \right|}
\newcommand{\ket}[1]{\left| #1 \right>}

\begin{document}

\title{Quantum communication with time-bin encoded microwave photons}

\author{P.~Kurpiers} \altaffiliation{P.K. and M.P. contributed equally to this work.} \affiliation{Department of Physics, ETH Z\"urich, CH-8093 Z\"urich, Switzerland}
\email[]{philipp.kurpiers@phys.ethz.ch, mpechal@stanford.edu, \\ andreas.wallraff@phys.ethz.ch}

\author{M.~Pechal} \altaffiliation{P.K. and M.P. contributed equally to this work.} \affiliation{Department of Applied Physics and Ginzton Laboratory, Stanford University, Stanford CA 94305 USA}

\author{B.~Royer}\affiliation{Institut Quantique and D\'epartment de Physique, Universit\'e de Sherbrooke, Sherbrooke, Qu\'ebec J1K 2R1, Canada}

\author{P.~Magnard} \affiliation{Department of Physics, ETH Z\"urich, CH-8093 Z\"urich, Switzerland}

\author{T.~Walter}\affiliation{Department of Physics, ETH Z\"urich, CH-8093 Z\"urich, Switzerland}

\author{J.~Heinsoo}\affiliation{Department of Physics, ETH Z\"urich, CH-8093 Z\"urich, Switzerland}

\author{Y.~Salath\'e}\affiliation{Department of Physics, ETH Z\"urich, CH-8093 Z\"urich, Switzerland}

\author{A.~Akin}\affiliation{Department of Physics, ETH Z\"urich, CH-8093 Z\"urich, Switzerland}

\author{S.~Storz}\affiliation{Department of Physics, ETH Z\"urich, CH-8093 Z\"urich, Switzerland}

\author{J.-C.~Besse}\affiliation{Department of Physics, ETH Z\"urich, CH-8093 Z\"urich, Switzerland}

\author{S.~Gasparinetti}\affiliation{Department of Physics, ETH Z\"urich, CH-8093 Z\"urich, Switzerland}

\author{A.~Blais}\affiliation{Institut Quantique and D\'epartment de Physique, Universit\'e de Sherbrooke, Sherbrooke, Qu\'ebec J1K 2R1, Canada}\affiliation{Canadian Institute for Advanced Research, Toronto, Canada}

\author{A.~Wallraff}\affiliation{Department of Physics, ETH Z\"urich, CH-8093 Z\"urich, Switzerland}

\date{\today}

\begin{abstract}
Heralding techniques are useful in quantum communication to circumvent losses without resorting to error correction schemes or quantum repeaters. Such techniques are realized, for example, by monitoring for photon loss at the receiving end of the quantum link while not disturbing the transmitted quantum state. We describe and experimentally benchmark a scheme that incorporates error detection in a quantum channel connecting two transmon qubits using traveling microwave photons. This is achieved by encoding the quantum information as a time-bin superposition of a single photon, which simultaneously realizes high communication rates and high fidelities. The presented scheme is straightforward to implement in circuit QED and is fully microwave-controlled, making it an interesting candidate for future modular quantum computing architectures.
\end{abstract}
\maketitle

Engineering of large-scale quantum systems will likely require coherent exchange of quantum states between distant units. The concept of quantum networks has been studied theoretically~\cite{Cirac1997,Jiang2007,Kimble2008,Northup2014} and substantial experimental efforts have been devoted 
to distribute entanglement over increasingly larger distances
~\cite{Chou2005,Moehring2007,Ma2012,Ritter2012,Roch2014,Hensen2015,Yin2017a,Delteil2017,Campagne-Ibarcq2018,Kurpiers2018,Axline2018}. In practice, quantum links inevitably experience losses, which vary significantly between different architectures and may range from $2\times 10^{-4}\,\mathrm{dB}/\mathrm{m}$ in optical fibers~\cite{Agrawal2011} to $5\times 10^{-3}\,\mathrm{dB}/\mathrm{m}$ in superconducting coaxial cables and waveguides at  cryogenic temperatures~\cite{Kurpiers2017}.
However, no matter which architecture is used, the losses over a sufficiently long link will eventually destroy the coherence of the transmitted quantum state, unless some measures are taken to mitigate these losses. Possible ways to protect the transmitted quantum information rely, for example, on using quantum repeaters~\cite{Briegel1998,Muralidharan2016}, error correcting schemes~\cite{Shor1995,Braunstein1998,Lloyd1998} or heralding protocols~\cite{Maunz2009,Hofmann2012,Usmani2012,Bernien2013}, which allow one to re-transmit the information in case photon loss is detected.

Heralding protocols are particularly appealing for near-term scaling of quantum systems since they are implementable without a significant resource overhead and can provide deterministic remote entanglement at predetermined times~\cite{Humphreys2018}. In essence, these protocols rely on encoding the transmitted quantum information in a suitably chosen subspace $S$ such that any error, which may be encountered during transmission, causes the system to leave this subspace. On the receiving end, a measurement which determines whether the system is in $S$ but does not distinguish between individual states within $S$, can be used to detect if an error occurred. Crucially, when the transfer is successful, this protocol does not disturb the transmitted quantum information. As a counter example, a simple encoding as a superposition of the vacuum state $|0\rangle$ and the single photon Fock state $|1\rangle$ is not suitable to detect errors due to photon loss because the error does not cause a transition out of the encoding subspace $\{|0\rangle,|1\rangle\}$. For this reason, encodings using other degrees of freedom such as polarization~\cite{Mattle1996,Peters2005}, angular momentum~\cite{MolinaTerriza2007,Yao2011}, frequency~\cite{Lukens2017}, time-bin~\cite{Brendel1999,Marcikic2004,Brecht2015b,Lee2018} or path~\cite{Kok2002,Matthews2011} are more common at optical frequencies. Heralding schemes have been used with superconducting circuits in the context of measurement based generation of remote entanglement~\cite{Roch2014,Dickel2018} and also using two-photon interference~\cite{Narla2016}. The more recent deterministic state transfer and remote entanglement protocols based on the exchange of shaped photon wave packets~\cite{Campagne-Ibarcq2018,Kurpiers2018,Axline2018,Leung2018} have to the best of our knowledge not yet been augmented using heralding protocols.

\begin{figure*}[t]
\begin{center}
\includegraphics{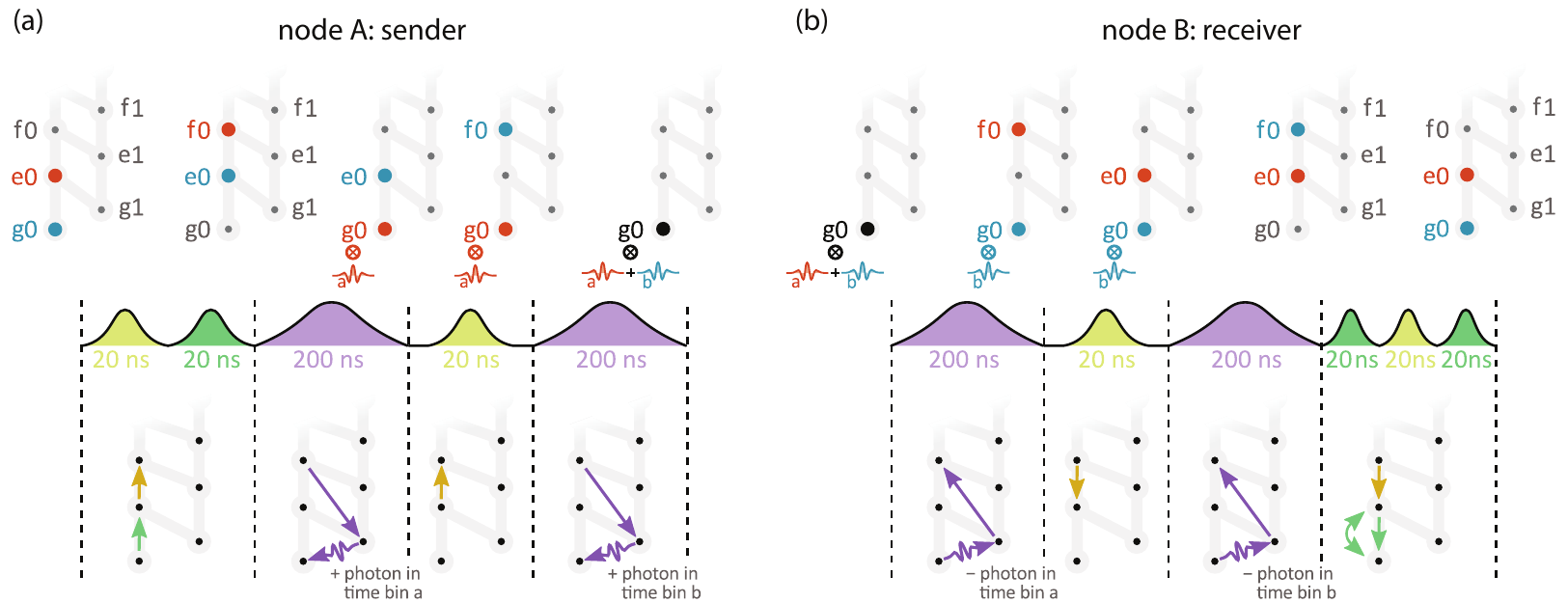}
\end{center}
\vspace{-0.2cm}
\caption{ Schematic representation of the pulse sequence implementing the time-bin encoding process at (a) the sender and (b) the receiver. In the level diagrams, the gray vertical lines represent Hamiltonian matrix elements due to microwave drive between transmon levels while the diagonal ones show the transmon-resonator coupling. (a) The qubit state is initially stored as a superposition of $|g0\rangle$ (blue dot) and $|e0\rangle$ (red dot). The first two pulses map this state to a superposition of $|e0\rangle$, $|f0\rangle$ and the third pulse transfers $|f0\rangle$ into $|g0\rangle$ while emitting a photon in time bin $a$ (see red symbol for photon in mode $a$). Then, after $|e0\rangle$ is swapped to $|f0\rangle$, the last pulse again transfers $|f0\rangle$ into $|g0\rangle$ and emits a photon in time bin $b$. (b) Reversing the protocol we reabsorb the photon, mapping the time-bin superposition back onto a superposition of transmon states.}
\vspace{-0.2cm}
\label{fig:timeBinEnc}
\end{figure*}

In this work, we propose and experimentally benchmark a method to transfer qubit states over a distance of approximately $0.9 \ \rm{m}$ using a time-bin superposition of two propagating temporal microwave modes. Our experimental results show that the protocol leads to a significant performance improvement, which is, in our case, a reduction of the transfer process infidelity by a factor of approximately two assuming ideal qubit readout. The stationary quantum nodes are transmons~\cite{Koch2007} coupled to coplanar waveguide resonators. The two lowest energy eigenstates of the transmon, $\ket{g}$ and $\ket{e}$, form the qubit subspace $S$ while the second excited state, $\ket{f}$, is used to detect potential errors. The multi-level nature of the transmon is also essential to the photon emission and reabsorption process, as described below. This technique can also be adapted to prepare entangled states of the qubit and the time-bin degree of freedom, making it suitable for heralded distribution of entanglement. Remarkably, it does not require any specialized components beyond the standard circuit QED setup with a transmon or any other type of non-linear multi-level system coupled to a resonator, such as capacitively shunted flux qubits~\cite{Yan2016}, and can be implemented without frequency tunability.

Our time-bin encoding scheme is based on a technique  for generating traveling microwave photons~\cite{Pechal2014} via a Raman-type transition in the transmon-resonator system~\cite{Zeytinoglu2015}. When the transmon is in its second excited state $|f\rangle$ and the resonator in the vacuum state $|0\rangle$, application of a strong microwave drive of appropriate frequency induces a second order transition from the initial state $|f0\rangle$ into the state $|g1\rangle$ where the transmon is in the ground state and the resonator contains a single photon. This photon is subsequently emitted into a waveguide coupled to the resonator, leaving
the transmon-resonator system in its joint ground state. As the magnitude and phase of the coupling between $|f0\rangle$ and $|g1\rangle$ is determined by the amplitude and phase of the applied drive~\cite{Pechal2014}, the waveform of the emitted photon can be controlled by shaping the drive pulse. The same process applied in reverse can then be used to reabsorb the traveling photon by another transmon-resonator system~\cite{Cirac1997,Kurpiers2018}.

The process for transferring quantum information stored in the transmon into a time-bin superposition state consists of the steps illustrated in Fig.~\ref{fig:timeBinEnc}(a): The transmon qubit at node A is initially prepared in a superposition of its ground and first excited state, $\alpha|g\rangle + \beta|e\rangle$, and the resonator in its vacuum state $|0\rangle$. Next, two pulses are applied to transform this superposition into $\alpha|e\rangle + \beta|f\rangle$. Then, another pulse induces the transition from $|f0\rangle$ to $|g1\rangle$ as described above, which is followed by spontaneous emission of a photon from the resonator. The shape of the $|f0\rangle$-$|g1\rangle$ drive pulse is chosen such that the photon is emitted into a time-symmetric mode centered around time $t_a$. After this first step, the system is in the state $\alpha|e0\rangle \otimes |0\rangle + \beta|g0\rangle\otimes |1_a\rangle$, where $|0\rangle$ and $|1_a\rangle$ denote the vacuum state of the waveguide and the single-photon state in the time-bin mode $a$. Next, the population from state $|e\rangle$ is swapped into $|f\rangle$ and the photon emission process is repeated, this time to create a single photon in a time-bin mode $b$ centered around time $t_b$. The resulting state of the system is $|g0\rangle \otimes \left(\alpha |1_b\rangle + \beta |1_a\rangle \right)$.

Because of time-reversal symmetry, a single photon, which is emitted by a transmon-resonator system into a propagating mode with a time-symmetric waveform, can be reabsorbed with high efficiency by another identical transmon-resonator system~\cite{Cirac1997}. This absorption process is induced by a drive pulse obtained by time-reversing the pulse that led to the emission of the photon. By reversing both drive pulses in the time-bin encoding scheme, as illustrated in Fig.~\ref{fig:timeBinEnc}(b), an incoming single photon in the time-bin superposition state $\alpha|1_b\rangle + \beta|1_a\rangle$ will cause the receiving transmon-resonator system, initialized in $|g0\rangle$, to be driven to the state $\alpha|e0\rangle + \beta|g0\rangle$ as the photon is absorbed. Thus, this protocol transfers the qubit state encoded as a superposition of $|g\rangle$ and $|e\rangle$ from transmon A to transmon B. In short, the sequence is
 \begin{align*}
  &\Big(\alpha |g\rangle_{A} + \beta |e\rangle_{A}\Big) \otimes |g\rangle_{B} \rightarrow\\ 
  &|g\rangle_{A} \otimes \Big(\alpha  |1_b\rangle + \beta |1_a\rangle\Big)
  \otimes |g\rangle_{B}\rightarrow \\
& |g\rangle_{A} \otimes \Big(\alpha |g\rangle_{B} + \beta |e\rangle_{B}\Big)
 \end{align*}
where we have omitted the states of the resonators and the propagating field whenever they are in their respective vacuum states. 

An important property of this transfer protocol is its ability to detect photon loss in the communication channel. Indeed, if a photon is lost or not absorbed by the receiver, system B receives a vacuum state at its input instead of the desired single-photon state. This means that both absorption pulse sequences will leave transmon B in its ground state $|g\rangle$ which will be subsequently mapped into $|f\rangle$ by the final three pulses. By performing a quantum non-demolition measurement on the transmon which distinguishes between $|f\rangle$ and the subspace spanned by $|g\rangle$,  $|e\rangle$, but does not measure within this subspace, we can detect the photon loss event without affecting the transmitted quantum information. Such a binary measurement of a qutrit state can, for example, be realized by suppressing the measurement-induced dephasing in the $ge$ subspace using parametric amplification and feedback~\cite{deLange2014} or by engineering the dispersive shifts of two transmon states on the readout resonator to be equal~\cite{Jerger2016a}. The protocol also detects failures of the state transfer due to energy relaxation at certain times during the time-bin encoding protocol, e.g.~if no photon is emitted from A due to decay to $\ket{g}$ before the first time bin. We discuss the detection of qutrit energy relaxation based on quantum trajectories in Appendix~\ref{app:MES}.

\begin{figure}[!ht] 
\centering
\includegraphics{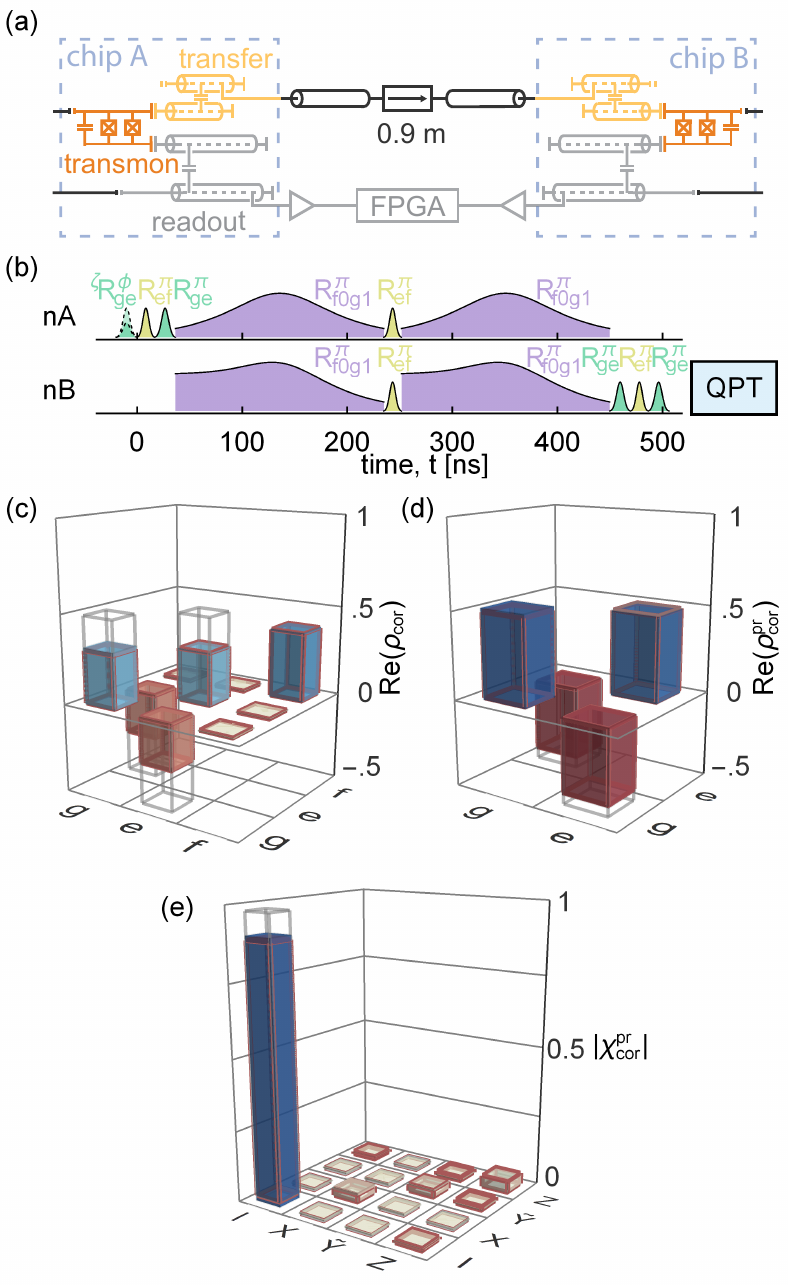}
\vspace{-0.1cm}
\caption{(a)~Simplified schematic of the circuit QED setup. At each of the nodes a transmon (orange) is capacitively coupled to two Purcell-filtered resonators which are used for readout (gray) and remote quantum communication (yellow). The directional quantum channel consists of a semi-rigid coaxial cable intersected by an isolator (see Ref.~\cite{Kurpiers2018} for details). (b)~Pulse scheme used to characterize the time-bin encoding protocol to transfer qubit states between two distant nodes using quantum process tomography (QPT). $^{\zeta}\rm{R}^{\phi}_{\rm{ij}}$ labels a Gaussian derivative removal by adiabatic gate (DRAG) microwave pulses~\cite{Motzoi2009,Gambetta2011} for transmon transition $\rm{ij} = \{\rm{ge},\,\rm{ef}\}$ or transmon-resonator transitions $\rm{ij} = \rm{f0g1}$ of angle $\phi = \{\pi/2,\,\pi\} $ around the rotation axis $\zeta = \{\rm{x},\rm{y}\}$. If no axis is specified the pulses are around the x-axis. (c)~Real part of the qutrit density matrix $\rho_{\rm{cor}}$ for the input state $\ket{-}=(\ket{g}-\ket{e})/\sqrt{2}$ after the state transfer protocol reconstructed using measurement-error correction. The magnitude of each imaginary part is $<0.017$. (d)~Projection of  $\rho_{\rm{cor}}$ (c) onto the $ge$ subspace $\rho_{\rm{cor}}^{\rm{pr}}$ performed numerically which we use to reconstruct the process matrix $\chi_{\rm{cor}}^{\rm{pr}}$ (absolute value shown in (e)). The colored bars show the measurement results, the gray wire frames the ideal density or process matrix, respectively. The results of numerical master equation simulations are depicted as red wire frames.}
\vspace{-0.3cm}
\label{fig:stateTransfer}
\end{figure}

We implemented this time-bin encoding protocol using the setup depicted in Fig.~\ref{fig:stateTransfer}(a)  (see Appendix~\ref{app:Parameters} and Ref.~\cite{Kurpiers2018} for details). We performed qutrit single-shot readout instead of the binary measurement at transmon B to characterize the quantum state transfer with process tomography. For that, we initialized both transmon qubits in their ground  states~\cite{Magnard2018,Egger2018} and subsequently prepared the qubit at node A in one of the six mutually unbiased qubit basis states $\ket{\psi_{\rm{in}}}=\{\ket{g},\ket{e},\ket{\pm}=(\ket{g}\pm\ket{e})/\sqrt{2},\ket{\pm i}=(\ket{g}\pm i\ket{e})/\sqrt{2}\}$~\cite{Enk2007}  [Fig.~\ref{fig:stateTransfer}(b)]. We then ran the time-bin encoding and reabsorption protocol, as described above [Fig.~\ref{fig:timeBinEnc}] and implemented quantum state tomography at node B for all six input states. Directly after the tomography pulses, we read out the $\ket{g}$, $\ket{e}$ and $\ket{f}$ states of transmon B with single-shot readout using a Josephson parametric amplifier (JPA)~\cite{Eichler2014a} in the output line. For readout characterization, we extracted probabilities of correct assignment of state $\ket{g}$, $\ket{e}$ and $\ket{f}$ for transmon B of $P_{\rm{g\ket{g}}}=P(g|\ket{g})=98.5 \%$, $P_{\rm{e\ket{e}}}=92.3 \%$ and $P_{\rm{f\ket{f}}}=86.4 \%$ (see also Appendix~\ref{app:QutritReadout}). Based on these single-shot measurements, we postselected experimental runs in which transmon B is not measured in the $\ket{f}$ state keeping on average $P_{\rm{suc}}^{qst}=64.6 \%$ of the data, and transferring qubit states at a rate $\Gamma_{\rm{qst}}/2\pi=P_{\rm{suc}}^{\rm{qst}}\Gamma_{\rm{exp}}/2\pi\approx  32.3 \, \rm{kHz}$. Using the post-selected data, we reconstructed the density matrices $\rho_{\rm{ps}}$ of the qubit output state at node B based solely on the single-shot readout results and obtain the process matrix $\chi_{\rm{ps}}$ of the quantum state transfer. We compute an averaged state fidelity of $F_{\rm{s}}^{\rm{ps}}=\rm{avg}(\braket{\psi_{\rm{in}}|\rho_{\rm{ps}}|\psi_{\rm{in}}})=88.2\pm 0.2\%$ and a process fidelity of $F_{\rm{p}}^{\rm{ps}}=\rm{tr}(\chi_{\rm{ps}}\chi_{\rm{ideal}})= 82.3 \pm 0.2 \%$ relative to the ideal input states $\ket{\psi_{\rm{in}}}$ and the ideal identity process, respectively.

To illustrate the detection of photon loss using the time-bin encoding protocol, we reconstructed all six qutrit density matrices $\rho_{cor}$ of the output state at node B using the same dataset and correcting for measurement errors in the qutrit subspace~\cite{Steffen2006a,Kurpiers2018} (see Appendix~\ref{app:QutritReadout} for details). These qutrit density matrices have a significant average population of level $\ket{f}$ of $39.1 \%$ indicating the detection of errors after the time-bin encoding protocol, which is compatible with $1-P_{\rm{suc}}^{qst}$ of the post-selected analysis [Fig.~\ref{fig:stateTransfer}(c)]. Next, we projected these density matrices numerically onto the qubit $ge$ subspace  $\rho_{cor}^{\rm{pr}}$ [Fig.~\ref{fig:stateTransfer}(d)], simulating an ideal error detection and reconstructed the process matrix $\chi_{\rm{cor}}^{\rm{pr}}$  of the quantum state transfer [Fig.~\ref{fig:stateTransfer}(e)].  
In this way, we found an average state fidelity of $F_{\rm{s}}^{\rm{cor}}=\rm{avg}(\braket{\psi_{\rm{in}}|\rho_{\rm{cor}}^{\rm{pr}}|\psi_{\rm{in}}})= 93.5\pm  0.1\%$ and a process fidelity of $F_{\rm{p}}^{\rm{cor}}=\rm{tr}(\chi_{\rm{cor}}^{\rm{pr}}\chi_{\rm{ideal}})=90.3 \pm 0.2 \%$ based on these measurement-corrected matrices. This analysis allowed us to compare the time-bin encoded protocol directly to a fully deterministic scheme without error mitigation, implemented in a similar setup~\cite{Kurpiers2018}, in which we obtained $F_{\rm{p}}^{\rm{det}} \approx 80 \%$. This clearly shows the advantage of time-bin encoding to reduce the effect of photon loss. In addition, we analyzed the sources of infidelity by performing numerical master equation simulations (MES) of the time-bin encoding protocol which we compared to the measurement-error corrected density and process matrices. We find excellent agreement with the experimental results, indicated by a small trace distance $\rm{tr}\left|\chi_{\rm{cor}}^{\rm{pr}}-\chi_\mathrm{sim}\right|/2=0.03$ which is ideally 0 for identical matrices and 1 for orthogonal ones. The MES results indicate that approximately $5.5 \%$ of the infidelity can be attributed to $\ket{f}\rightarrow \ket{e}$ and $\ket{e}\rightarrow \ket{g}$ energy relaxation at both transmons during the protocol. Pure qutrit dephasing can explain the remaining infidelity.

\begin{figure}
\centering
\includegraphics{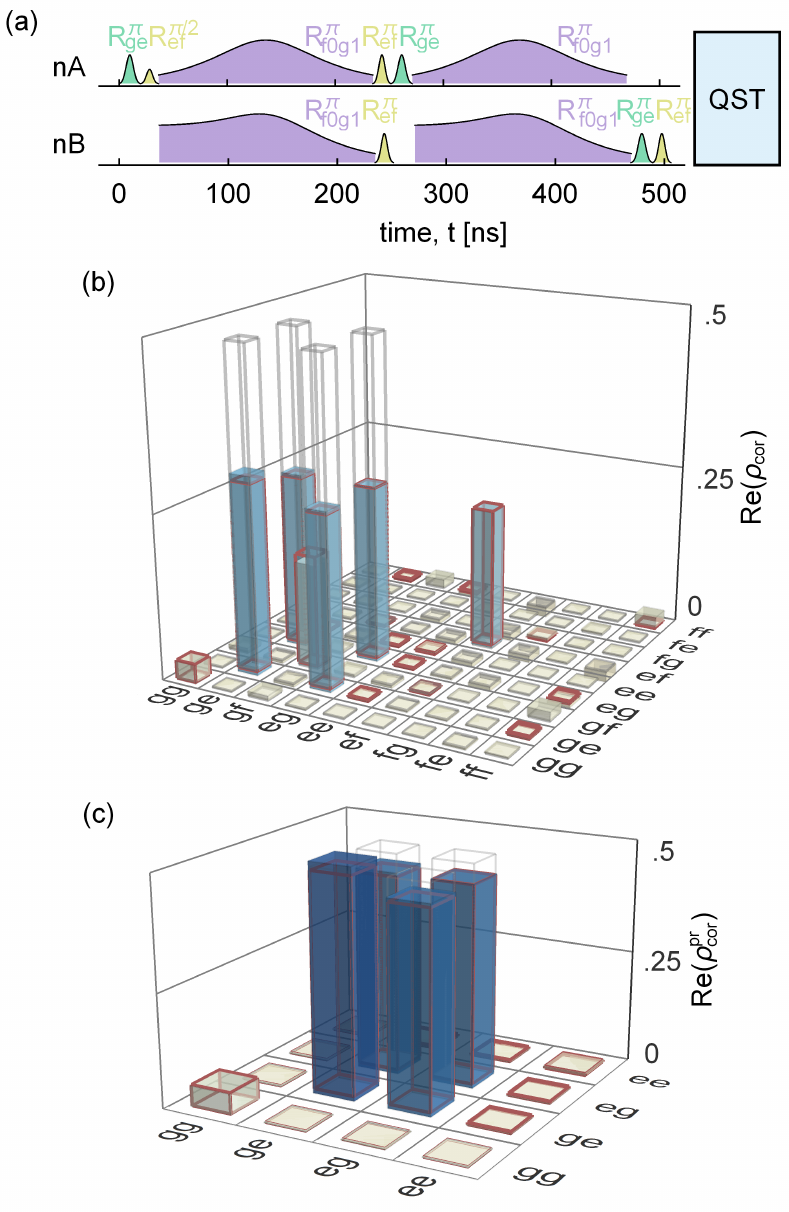}
\vspace{-0.1cm}
\caption{(a)~Pulse scheme for generating remote entanglement between nodes A and B (see text for details). (b)~Real part of the two-qutrit density matrix $\rho_{\rm{cor}}$ reconstructed after execution of the time-bin remote entanglement protocol using measurement-error correction. The colored bars indicate the measurement results, the ideal expectation values for the Bell state $\ket{\psi^+}=(\ket{g_A,e_B}+\ket{e_A,g_B})/\sqrt{2}$ are shown as gray wire frames and the results of a master equation simulation as red wire frames.  The magnitude of the imaginary part of each element of the density matrix is $<0.024$. (c)~Numerical projection of the $\rho_{\rm{cor}}$ onto the $ge$ subspace to obtain the two-qubit density matrix $\rho_{\rm{cor}}^{\rm{pr}}$ which is in excellent agreement with our MES (trace distance of $0.028$).}
\vspace{-0.1cm}
\label{fig:BellState}
\end{figure}

In addition to a direct quantum state transfer, the generation  of entanglement between distant nodes is a key task of quantum communication. Here, we use a simple modification of the state-transfer protocol to perform this task [Fig.~\ref{fig:BellState}(a)]. Both transmon-resonator systems are first initialized in their ground states. The first two pulses of the remote-entanglement protocol prepare transmon A in an equal superposition state $1/\sqrt{2}(|e\rangle+|f\rangle)$, followed by a pulse sequence which entangles the transmon state $\ket{g}$ and $\ket{e}$ with the time-bin qubit and maps the state of the time-bin qubit to transmon B. This process can be summarized as
\begin{align*}
 & \frac{1}{\sqrt{2}}\Big((|e\rangle_{A}+|f\rangle_{A}\Big) \otimes |g\rangle_{B} \rightarrow\\
  & \frac{1}{\sqrt{2}}\Big( |g\rangle_{A}\otimes|1_a\rangle + |e\rangle_{A}\otimes |1_b\rangle\Big) \otimes |g\rangle_{B} \rightarrow\\
 & \frac{1}{\sqrt{2}}\Big( |g\rangle_{A}\otimes|e\rangle_{B} + |e\rangle_{A}\otimes|g\rangle_{B}\Big)
\end{align*}
and, in case of an error, transmon B ends up in state $|f\rangle_{B}$.

We performed post-selected experiments of the entanglement-generation protocol by selecting only experimental runs in which neither qutrit was measured in the $\ket{f}$ state using individual single-shot readout on both transmons. Under this condition, we retained $P_{\rm{suc}}^{\rm{ent}}\approx 61.5 \%$ of the data and obtained a Bell state fidelity of $F_{\rm{s}}^{\rm{ps}}=\braket{\psi^+|\rho_{\rm{ps}}|\psi^+} = 82.3 \pm 0.4 \%$ compared to an ideal Bell state $\ket{\psi^+}=(\ket{g_A,e_B}+\ket{e_A,g_B})/\sqrt{2}$. In these post-selected experiments, we generated entangled states at rate $\Gamma_{\rm{ent}}/2\pi=P_{\rm{suc}}^{\rm{ent}}\Gamma_{\rm{exp}}/2\pi\approx 30.8 \, \rm{kHz}$. To benchmark this entanglement protocol, we used full two-qutrit state tomography of the transmons in which we corrected for measurement errors with the same data set. The reconstructed density matrix, shown in Fig.~\ref{fig:BellState}(c), displays a high population of the $\ket{g_A,f_B}$, $\ket{e_A,f_B}$ states, $P_{gf}=16.0\%$ and $P_{ef}=21.4\%$, and small population of $\ket{f_A,g_B}$, $\ket{f_A,e_B}$ and $\ket{f_A,f_B}$ ,$\sum_{i=\{g,e,f\}} P_{fi}=2.7\%$, which indicates that photon loss is a significant source of error. We projected onto the $ge$ qubit subspace numerically and obtained a two-qubit density matrix, Fig.~\ref{fig:BellState}(d), showing a fidelity of $F_{\rm{s}}^{\rm{cor}}=\braket{\psi^+|\rho_{\rm{cor}}^{\rm{pr}}|\psi^+}=92.4 \pm 0.4 \%$. Comparing this state fidelity to the fully deterministic case, $F_{\rm{s}}^{\rm{det}}\approx 79 \%$~\cite{Kurpiers2018}, shows the potential of the proposed time-bin encoding protocol to generate remote  entanglement independent of photon loss. Using a MES we attribute approximately $6.5\%$ of the infidelity to energy relaxation and the rest to dephasing. As detailed in Appendix~\ref{app:MES}, we performed a MES based on quantum trajectories and find that $64 \%$ of all decay events during the time-bin encoding protocol are detected. However, due to the additional time needed for performing the time-bin encoding protocol relative to the direct Fock-state encoding this protocol is affected more by pure qutrit dephasing.

In conclusion, we experimentally demonstrated a method for transferring a qubit state between a three-level superconducting quantum circuit and a time-bin superposition of a single propagating microwave photon. This type of encoding lends itself naturally to quantum communication protocols which allow detection of photon loss in the quantum link while maintaining a high communication rate. In our experiment, we have observed that the described protocol significantly improves the fidelity of transmitted quantum states and distributed Bell states  between two distant transmon qubits when outcomes are post-selected on successful transmission of a photon. We also observed and analyzed the potential of the time-bin encoding protocol to detect errors due to energy relaxation of the qutrits during the protocol.

\section*{Acknowledgements}
This work was supported by the European Research Council (ERC) through the 'Superconducting Quantum Networks' (SuperQuNet) project, by the National Centre of Competence in Research 'Quantum Science and Technology' (NCCR QSIT) a research instrument of the Swiss National Science Foundation (SNSF), by Baugarten Stiftung, by ETH Zürich Foundation, by ETH Zürich, by NSERC, the Canada First Research Excellence Fund, and by the Vanier Canada Graduate Scholarships. Marek Pechal was supported by the Early Postdoc Mobility Fellowship of the Swiss National Science Foundation (SNSF), and by the National Science Foundation (NSF) grant ECCS-1708734.

\section*{Author contributions}
The time-bin encoding protocol was developed by M.P. and P.K. The experimental implementation was designed by P.K., P.M., T.W and M.P. The samples were fabricated by J.-C.B., T.W. and S.G. The experiments were performed by P.K. The data was analyzed and interpreted by P.K., M.P., B.R., A.B. and A.W. The FPGA firmware and experiment automation was implemented by J.H., Y.S., A.A., S.S, P.M. and P.K. The master equation simulation were performed by B.R., P.K and M.P. The manuscript was written by M.P., P.K., B.R., A.B. and A.W. All authors commented on the manuscript. The project was led by A.W.

\appendix

\section{Stochastic Master Equation Simulations}
\label{app:MES}
\begin{figure}[t]
\centering
\includegraphics{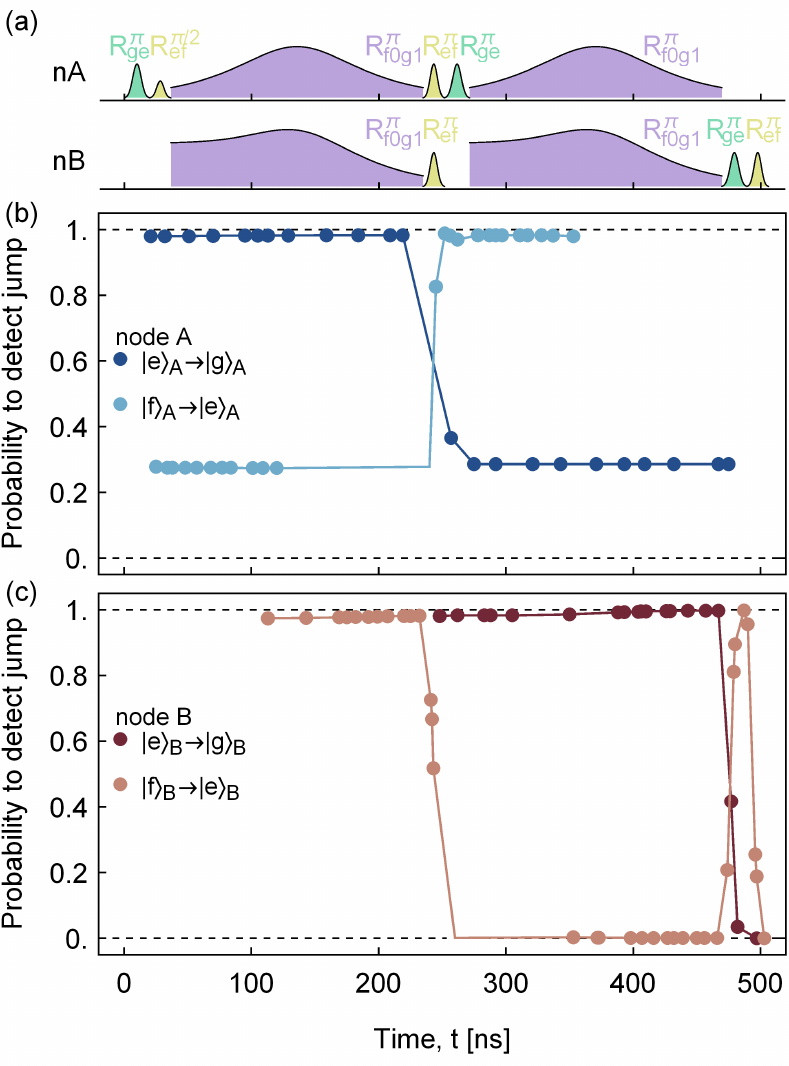}
\vspace{-0.1cm}
\caption{(a)~Pulse scheme for generating remote entanglement between node A and B. Probability that a quantum jump from $\ket{f}\rightarrow \ket{e}$ or $\ket{e}\rightarrow \ket{g}$ of the qutrit at node A (b) or node B~(c) is detected during the time-bin remote entanglement protocol. The data is based on 2000 stochastic trajectories of a numerical master equation simulation (see Appendix~\ref{app:MES} for details).}
\label{fig:MES}
\end{figure}

In order to investigate the robustness of the time-bin encoding protocol with respect to qutrit energy relaxation, we numerically study a fictitious experiment with the same device parameters in which we monitor quantum jumps from $\ket{f}\rightarrow \ket{e}$ and from $\ket{e}\rightarrow \ket{g}$ on qutrits A and B, conditioning the system state on the measurement record.
We perform $N_{\mathrm{traj}} = 2000$ trajectories unraveled in the manner described above~\cite{Breuer2002} and, for each realization for which a jump occurred, compute the probability that the error is detected at the end of the time-bin protocol. Fig.~\ref{fig:MES}(b) and (c) show this heralding probability as a function of the time at which the jump occurred for the four energy relaxation processes, $\ket{f}_{\rm{n}}\rightarrow \ket{e}_{\rm{n}}$ and $\ket{e}_{\rm{n}}\rightarrow \ket{g}_{\rm{n}}$, $\rm{n}=\{\rm{A},\rm{B}\}$.   
For some decay events, qutrit B ends in the $\ket{f}$ state and, consequently, the time-bin encoding protocol partially allows the heralding of qutrit relaxation.

For simplicity, Fig.~\ref{fig:MES}(b) and (c) only show the cases where a single decay event occurred. Summing over all trajectories where one or more jumps are present, we can estimate the probability of an undetected decay event,
\begin{equation*}
P_{\rm{undet}} = \frac{1}{N_{\mathrm{traj}}}\sum_{\mathrm{jump}} \left( 1 - \mathrm{Tr}\left[\ketbra{f}_B \rho_{\mathrm{jump}}(t_f)\right]\right),
\end{equation*}
where $\rho_{\mathrm{jump}}(t_f)$ is the state conditioned on the measurement record at the end of the protocol. For the entanglement generation protocol and the parameters of this experiment, we find that the probability of an undetected decay event is $P_{\rm{undet}}=5.5\%$ which corresponds to $P_{\rm{undet}}/P_{\rm{jump}}=35.7\%$ of trajectories where a jump occurred. Three types of decay events contribute the most to $P_{\rm{undet}}$, as shown in Fig.~\ref{fig:MES}(b) and (c). At node A, $\ket{f}_A \rightarrow \ket{e}_A$ jumps (light blue points) in the first time bin are generally not detected since this results in a photon being emitted in the second time bin, with qutrit B ending in $\ket{e}$. However, if there is also a photon loss event in the communication channel in the second time bin, then qutrit B ends in $\ket{f}$ and the combination of the two errors is detected. When the communication channel is perfect, this type of decay event is not detected. Second, $\ket{e}_A \rightarrow \ket{g}_A$ jumps (dark blue points) in the second time bin are generally not detected and there are two ways they can occur. If there are no other errors in the protocol, then qutrit B ends in $\ket{g}$ and the error is not detected. However, a $\ket{e}_A \rightarrow \ket{g}_A$ jump can also occur if there is a photon loss event in the first time bin. In that situation, qutrit B ends $\ket{f}$ and the error is detected. At node B,
$\ket{f}_B \rightarrow \ket{e}_B$ jumps (beige points) in the second time bin are not detected since qutrit B ends in $\ket{g}$.

In contrast to energy relaxation, pure dephasing does not lead to direct changes in qutrit populations. Consequently, the time-bin encoding does not allow the heralding of phase errors.


\section{Sample and Setup}
\label{app:Parameters}
The samples and setup are identical to the one of Ref.~\cite{Kurpiers2018}.  Up to an exchange of the cryogenic coaxial circulator (Raditek RADC-8-12-Cryo) in the connection between the two samples with a rectangular waveguide isolator (RADI-8.3-8.4-Cryo-WR90) which affected the bandwidth of the transfer resonators due to its different impedance. The device parameters are summarized in Table~\ref{tab:ParameterSummary}.

\begin{table} [t]
\footnotesize
\begin{tabular}{|l|rr|}
\hline 
quantity, symbol (unit) & \hfill{}Node A\hfill{} & \hfill{}Node B\hfill{}\tabularnewline
\hline 
readout resonator frequency, $\nu_{\mathrm{R}}$~(GHz)                  & 4.788  & 4.780\tabularnewline
readout Purcell filter frequency, $\nu_{\mathrm{Rpf}}$~(GHz)           & 4.778  & 4.780\tabularnewline
readout resonator bandwidth, $\kappa_{R}/2\pi$~(MHz)                   & 12.6   & 27.1\tabularnewline
readout circuit dispersive shift, $\chi_{R}/2\pi$~(MHz)                & 5.8    & 11.6\tabularnewline
transfer resonator frequency, $\nu_{\mathrm{T}}$~(GHz)                 & 8.400 & 8.400\tabularnewline
transfer Purcell filter frequency, $\nu_{\mathrm{Tpf}}$~(GHz)          & 8.426  & 8.415\tabularnewline
transfer resonator bandwidth, $\kappa_{T}/2\pi$~(MHz)                  & 7.4   & 12.6\tabularnewline
transfer circuit dispersive shift, $\chi_{T}/2\pi$~(MHz)               & 6.3    & 4.7\tabularnewline
qubit transition frequency, $\nu_{\mathrm{ge}}$~(GHz)                  & 6.343  & 6.098\tabularnewline
transmon anharmonicity, $\alpha$~(MHz)                                 &-265    &-306\tabularnewline
$\ket{f,0} \leftrightarrow\ket{g,1}$ transition frequency, $\nu_{\mathrm{f0g1}}$~(GHz) &  4.021  & 3.490 \tabularnewline
$\ket{f,0} \leftrightarrow\ket{g,1}$ max. eff. coupling, $\tilde{g}_{\rm{m}}/2\pi$ ~(MHz)       &  5.4    &  5.6  \tabularnewline  
energy relaxation time on $ge$, $T_{\mathrm{1ge}}$~($\mathrm{\mu s}$)  & 5.0$\pm$0.4    & 4.5$\pm$0.2\tabularnewline
energy relaxation time on $ef$, $T_{\mathrm{1ef}}$~($\mathrm{\mu s}$)  & 2.3$\pm$0.4    & 1.6$\pm$0.1\tabularnewline
coherence time on $ge$, $T^{\rm{R}}_{\mathrm{2ge}}$~($\mathrm{\mu s}$) & 7.7$\pm$0.5    & 5.7$\pm$0.2\tabularnewline
coherence time on $ef$, $T^{\rm{R}}_{\mathrm{2ef}}$~($\mathrm{\mu s}$) & 3.2$\pm$1.3    & 2.3$\pm$0.7\tabularnewline
\hline 
\end{tabular}
\caption{\label{tab:ParameterSummary} Device parameters for nodes A and B.}
\end{table}

\begin{table*}
\footnotesize
\begin{centering}
\begin{tabular}{|c|ccccccccc|}
\hline 
&gg & ge & gf  & eg & ee & ef & fg & fe &ff \tabularnewline
\hline 
gg&0.027 &-0.004$i$ & 0.007-0.007$i$ &-0.001-0.011$i$ & 0 &-0.001$i$ & 0 &-0.003 & 0.002+0.002$i$\tabularnewline
ge&0.004$i$& 0.301&-0.005-0.024$i$& 0.269& 0.004$i$&-0.003+0.007$i$ &-0.002-0.006$i$ &-0.001 &-0.001\tabularnewline
gf&.007+0.007$i$&-0.005+0.024$i$& 0.16& 0.004+0.013$i$&-0.005+0.003$i$&-0.002-0.001$i$& 0.001-0.002$i$& 0.001+0.016$i$& 0.014-0.004$i$\tabularnewline
eg&-0.001+0.011$i$& 0.269& 0.004-0.013$i$& 0.268&-0.001-0.001$i$&-0.003+0.008$i$&-0.007-0.010$i$&-0.001$i$&-0.006-0.004$i$\tabularnewline
ee&0.&-0.004$i$&-0.005-0.003$i$&-0.001+0.001$i$& 0.003&-0.008+0.011$i$&-0.003$i$& 0.& 0.001\tabularnewline
ef&0.001$i$&-0.003-0.007$i$&-0.002+0.001$i$&-0.003-0.008$i$&-0.008-0.011$i$& 0.214&-0.011+0.017$i$& 0.001-0.001$i$&-0.016-0.007$i$\tabularnewline
fg&0.&-0.002+0.006$i$& 0.001+0.002$i$&-0.007+0.01$i$& 0.003$i$&-0.011-0.017$i$& 0.006&-0.001+0.001$i$& 0.001+0.001$i$\tabularnewline
fe&-0.003&-0.001& 0.001-0.016$i$& 0.001$i$& 0.& 0.001+0.001$i$&-0.001-0.001$i$& 0.003& 0.\tabularnewline
ff&0.002-0.002$i$&-0.001& 0.014+0.004$i$&-0.006+0.004$i$& 0.001&-0.016+0.007$i$& 0.001-0.001$i$& 0.& 0.018\tabularnewline
\hline 
\end{tabular}
\par\end{centering}
\caption{\label{tab:BellState} Numerical values of the experimentally obtained density matrix elements of the two-transmon remote entangled state in a two-qutrit basis using the time-bin encoding. The real part of this density matrix is depicted as colored bars in Fig.~\ref{fig:BellState}(b).}
\end{table*}

\section{Qutrit single-shot readout and population extraction}
\label{app:QutritReadout}
We estimated the measurement assignment probabilities $P_{\rm{s'\ket{s}}}=P(s'|\ket{s})$ by first assigning each trace prepared in state $\ket{s}$ to state $s'$ obtained from a single-shot measurement and normalized the recorded counts. We summarized those normalized counts in a vector $N_{\rm{B}}^{i}$ for each measured trace $i$. To obtain the assignment probabilities matrix $R_{\rm{A}}=P_{\rm{A}}(s_{\rm{A}}'|\ket{s_{\rm{A}}})=(N_{\rm{A}}^{\ket{g}},N_{\rm{A}}^{\ket{e}},N_{\rm{A}}^{\ket{f}})$ and $R_{\rm{B}}=P_{\rm{B}}(s_{\rm{B}}'|\ket{s_{\rm{B}}}$ for transmon A and B, we reset both transmons to their ground state, prepared them in either $\ket{g}$, $\ket{e}$ or $\ket{f}$ individually using DRAG microwave pulses and performed single shot readout for which we optimized the readout power and integration time~\cite{Walter2017} to minimize the sum of all measurement misidentifications (the off-diagonal elements of $R$)~\cite{Kurpiers2018}. For the single-shot readout we used Josephson parametric amplifiers (JPAs) with a gains of $21 \, \rm{dB}$ and $24 \, \rm{dB}$ and bandwidths of $20 \, \rm{MHz}$ and $ 28 \, \rm{MHz}$. We obtained the assignment probabilities matrix $R_{\rm{A}}=$
\begin{equation*}
\begin{array}{c | c c c}
	    & \ket{g}  & \ket{e} & \ket{f}  \\
        \hline
	g   & 97.8     & 2.7     & 2.9      \\
	e   & 0.7      & 93.8    & 3.7       \\
	f   & 1.5      & 3.5     & 93.4       \\
\end{array}
\end{equation*}
for transmon A for a readout time of $t_{\rm{r}}^{\rm{A}}=96 \, \rm{ns}$ and a state-dependent number of photons in the readout resonator $n_{\rm{r}}^{\rm{A}}$ of $0.5$ to $2$. For transmon B we computed $R_{\rm{B}}=$
\begin{equation*}
\begin{array}{c | c c c}
	    & \ket{g}  & \ket{e} & \ket{f}  \\
        \hline
	g   & 98.5     & 3.8     & 1.1      \\
	e   & 0.9      & 92.3    & 12.5       \\
	f   & 0.6      & 3.9     & 86.4       \\
\end{array}
\end{equation*}
for $t_{\rm{r}}^{\rm{B}}=216 \, \rm{ns}$ and $n_{\rm{r}}^{\rm{B}}$ between $0.2$ and $0.5$ used for characterizing the quantum state transfer protocol.

For two-qutrit states the assignment probability matrix $R_{\rm{AB}}=P_{\rm{AB}}(s_{\rm{A}}',s_{\rm{B}}'|\ket{s_{\rm{A}},s_{\rm{B}}})=R_{\rm{A}}R_{\rm{B}}$ can be calculated using the outer product of the single-qutrit assignment probabilities matrices.

In the post-selected measurement analysis we discarded traces which were assigned to $s'=f$ in the single-shot measurement, keeping only the $ge$ qubit subspace. We normalized the $g$, $e$ counts and set the normalized counts equal to the populations of transmons in the qubit subspace. Based on these population we reconstructed the density matrices of the output states after the state-transfer or entanglement-generation protocol using a maximum-likelihood approach~\cite{Smolin2012}.

To reconstruct the full single-qutrit (two-qutrit) density matrices we inverted the assignment probability matrix $R_{\rm{B}}^{-1}$ ($R_{\rm{AB}}^{-1}$) and obtained the qutrit population for each measured trace $M_{\rm{B}}^{i}=R_{\rm{B}}^{-1}\,N_{\rm{B}}^{i}$  (two-qutrit correlations $C_{\rm{AB}}=R_{\rm{AB}}^{-1}\,N_{\rm{AB}}$). Applying $R_{\rm{B}}^{-1}$ ($R_{\rm{AB}}^{-1}$) is an advantage since we can prepare qutrit states with a higher fidelity than performing qutrit readout~\cite{Kurpiers2018} because of an efficient reset~\cite{Magnard2018} and high fidelity single qutrit pulses.

\begin{table}[h]
\footnotesize
\begin{centering}
\begin{tabular}{|c|cccc|}
\hline 
&I & X &$\tilde{\rm{Y}}$ & Z \tabularnewline
\hline 
I                & 0.903 & -0.003-0.002$i$ &  -0.004& 0.007+0.005$i$  \tabularnewline
X                & -0.003+0.002$i$& 0.033& -0.007& 0.001   \tabularnewline
$\tilde{\rm{Y}}$ & -0.004& -0.007& 0.027& -0.003+0.002$i$ \tabularnewline
Z                &  0.007-0.005$i$& 0.001& -0.003-0.002$i$& 0.037      \tabularnewline
\hline 
\end{tabular}
\par\end{centering}
\caption{\label{tab:stateTransfer} Numerical values of the experimentally obtained process matrix elements of the qubit state transfer using the time-bin encoding protocol. The absolute value of this process matrix is depicted in Fig.~\ref{fig:stateTransfer}(e) as colored bars.}
\end{table}

\section{Measurement data}
\label{app:MeasData}
The measurement results of the quantum process tomography used to characterize the state-transfer protocol and of the two-qutrit density matrix after the remote entanglement protocol are shown in Table~\ref{tab:BellState} and Table~\ref{tab:stateTransfer}.

\bibliography{QudevRefDB}

\end{document}